# Appraising the absolute limits of nanotubes/nanospheres to preserve high-pressure materials

Yin L. Xu[1,*], Guang F. Yang[1,*], Yi Sun[1], Hong X. Song[1], Yu S. Huang[1], Hao Wang[1,†], Xiao Z. Yan[1,2,†] and Hua Y. Geng[1,3,†]

[1] *National Key Laboratory of Shock Wave and Detonation Physics, Institute of Fluid Physics, China Academy of Engineering Physics, Mianyang, Sichuan 621900, P. R. China;*

[2] *School of Science, Key Laboratory of Low Dimensional Quantum Materials and Sensor Devices of Jiangxi Education Institutes, Jiangxi University of Science and Technology, Ganzhou 341000, Jiangxi, P. R. China;*

[3] *HEDPS, Center for Applied Physics and Technology, and College of Engineering, Peking University, Beijing 100871, P. R. China.*

**Abstract:** Matter under high pressure often exhibits attractive properties, which, unfortunately, are typically irretrievable when released to ambient conditions. Intuitively, nanostructure engineering might provide a promising route to contain high-pressure phase of materials because of the exceptional mechanical strength at nanoscale. However, there is no available theoretical model that can analyze this possibility, not to mention to quantitatively evaluate the pressure-bearing capability of nano-cavities. Here, a physical model is proposed to appraise the absolute theoretical limit of various nanotubes/nanospheres to preserve high-pressure materials to ambient conditions. By incorporating with first-principles calculations, we screen and select four types of representative nanomaterials: graphene, hexagonal boron nitride (h-BN), biphenylene, and $\gamma$-graphyne, and perform systematic investigations. The results indicate that nanotube/nanosphere of graphene exhibits the best pressure-bearing capability, followed by h-BN, biphenylene and $\gamma$-graphyne. Our model reveals that the structure with the largest average binding energy per bond and the highest density of bonds will have the highest absolute limit to contain pressure materials, while electron/hole doping and interlayer interactions have minor effects. Our finding suggests that one can utilize nanotube/nanosphere with multiple layers to retrieve compressed material with higher pressures. For example, a single layer graphene sphere can retrieve compressed $LaH_{10}$ with a volume size of 26 $nm^3$ that corresponding to a pressure of 170 GPa and with a near room temperature superconductor transition of $T_c$=250 K. Similarly, in order to retrieve the metastable atomic hydrogen or molecular metallic hydrogen at about 250 GPa, it requires only three layers of a nanosphere to contain a volume size of 173 $nm^3$.

\* *Contributed equally.*
† *Corresponding authors. E-mail: wh_95@qq.com; yanxiaozhen@jxust.edu.cn; s102genghy@caep.cn*

# I. Introduction

Pressure plays a critical role in determining the structure and properties of materials[1]. Under extreme compression conditions, materials exhibit exotic but intriguing characteristics compared to their ambient-pressure states, including structural phase transitions[2], electronic orbital hybridizations[3], reordering of band structures[4] etc., which could lead to exceptional phenomena such as metal-nonmetal transition in electrides[5], collapse of magnetic moment[6], cold melting (LiNa[7] and H[8]), mobile solid state[9] and simultaneous superconducting superfluid state (H[10]), near room temperature superconductors (H and hydrides[10,11]), reactive inert elements[5,12], and highly energetic single-bonded polymeric nitrogen[13], just to mention a few. These high-pressure phases not only provide deeper comprehension of fundamental physics and chemistry under extreme conditions, but also demonstrate remarkable promise of compressed materials in superconducting and energy storage applications[14-17].

In particular, extreme compression provides an effective approach to tune and break through the material property limits, like high superconducting transition temperature ($T_c$)[10,18,19]. But the challenge to retrieve these high-pressure phases to ambient conditions is obstructing its practical applications. For example, metallic hydrogen is predicted to be a superconductor with a transition temperature $T_c$ higher than 300 K, but it will become spontaneously unstable when pressure is below 250 GPa at room temperature[20,21]. Similarly, hydrogen-rich compounds (e.g., $H_3S$[22,23] and $LaH_{10}$[24-26]) demonstrate remarkable $T_c$ exceeding 200 K under 150-200 GPa, but such property is lost when pressure is released. In order to lower the pressure threshold, researchers have tried various strategies. For example, introducing ternary elements, such as the La-B-H and La-Be-H systems[27,28], can achieve a superconducting transition temperature of 191 K and 156 K at 50 and 55 GPa, respectively. But it seems very unlikely to go further to lower the pressure to have a $T_c$ near or above room temperature. The difficult situation is that these properties can be applied only when they are retrievable. Unfortunately, most such phases cannot be naturally retrieved.

In principle, requiring natural metastability is not the only possible route to recover compressed materials. Imposing external confinement is another interesting alternative to contain high pressure materials. For example, Zeng *et al.*[29] developed nanostructured diamond capsules (NDCs) that enables preservation of substances in high-pressure states (up to 22 GPa) without requiring traditional high-pressure devices support. This strategy can go further to reduce the supporting material by incorporating with nano-cavity engineering, even via self-assembly. It is well-known that nanomaterials could be engineered to have unique mechanical properties, such as exceptional tensile strength[30] and large cavity volumes[31,32], and can be formed as nanotubes/nanospheres to encapsulate and retrieve high-pressure state materials. Unfortunately, currently there is no available theoretical model that can estimate the utmost theoretical limit of ideal nanostructures to contain high-pressure materials, making experimental efforts toward this goal lacking fundamental physical basis and theoretical stimulus, not to mention to optimize the experimental design to overcome the synthesis challenges.

In order to lay down a foundation for this direction, in this work, we propose a physical model to estimate the theoretical upper limit (i.e., the absolute limit) of nanotube/nanosphere to assess their capability to preserve high-pressure compressed materials when being engineered into a nanocavity. Various 2D nanosheets are explored and screened, and only the most promising four representative nanomaterials are selected for systematical investigations: graphene, h-BN, biphenylene, and $\gamma$-graphyne. Within them, first-principles calculations indicate that the graphene nanotubes/nanospheres have the highest pressure-bearing capability, followed by h-BN, biphenylene, and $\gamma$-graphyne. This ordering fulfils a rule that the structure with the largest average binding energy per bond and the highest bond density is the best for pressure preservation. We also find that chemical doping and interlayer interactions have insignificant effects on this issue, implying that adding more layers to the nanotubes/nanospheres can effectively enhance their pressure-bearing capability.

## II. Theory and Computational Method

## A. Modelling

For brevity, graphene nanosheet is employed to illustrate the physical model for preserving high-pressure materials. There is a unique edge state along the armchair direction of graphene (see the red part in Fig. 1a). It, as any other 2D nanosheets, can be rolled into nanotube or nanosphere as shown in Fig. 1(b). The formed nano-cavity can be utilized to store high-pressure materials. In this configuration, the outermost part represents the nanotube that provides the binding forces, while the inside region contains the material to be preserved, as shown in Fig. 1(c).

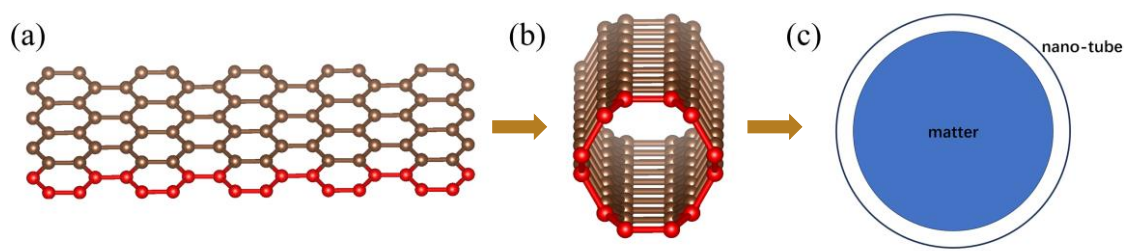

**Fig. 1** (color online) (a) Schematic diagram of the graphene structure, where the red atoms represent the edge states along the armchair direction; (b) Graphene rolled into a nanotube along the armchair direction; (c) A cross-sectional view of the nanotube containing high-pressure matter.

**(A.I) Monolayer nanotube**

A physical model of monolayer nanotube as sketched in Fig. 1(c) that describes the utmost limit of the capability to preserve high-pressure matter can be constructed as follows. Assuming that the monolayer nanotube has a radius of $r$ and a length of $l$, the total energy of the system can be written as:

$$E = E_m + E_n \tag{1}$$

where $E_m$ represents the internal energy of the contained material, and $E_n$ represents the energy of the nanotube itself, with

$$E_n = \epsilon \cdot S = \epsilon \cdot 2\pi r l + \epsilon \cdot 2\pi r^2 \tag{2}$$

$\epsilon$ denotes the energy density of the nanotube per unit area. The second term in Eq. (2) comes from the energy contribution by sealing the two side-terminals of the nanotube. Here we ignore the energy due to matter-nanotube interaction, by assuming this contribution is invariant during the system in dilation or shrinking process, thus can be eliminated in total energy variation. The variation of the total system energy with respect to volume change is:

$$\frac{\delta E}{\delta V} = \frac{\delta E_m}{\delta V} + \frac{\delta E_n}{\delta V} = \frac{\delta E_m}{\delta V} + \frac{\delta E_n}{\delta r} \cdot \frac{\delta r}{\delta V} + \frac{\delta E_n}{\delta l} \cdot \frac{\delta l}{\delta V} \tag{3}$$

The stationary condition of the system requires $\frac{\delta E}{\delta V} = 0$, we thus have:

$$\frac{\delta E_m}{\delta V} = -\frac{\delta E_n}{\delta V} = -(\frac{\delta E_n}{\delta r} \cdot \frac{\delta r}{\delta V} + \frac{\delta E_n}{\delta l} \cdot \frac{\delta l}{\delta V}) = -P \quad (4)$$

Assuming that the nanotube has large enough primitive cells, which is an appropriate approximation for the purpose of here to look for a theoretical upper limit of pressure preserving, we consider a case that there are $n$ primitive cells along the circumference direction with a lattice constant of $a$, and $m$ primitive cells in the axial direction with a lattice constant of $b$. When the radius of nanotube is sufficiently large, they can be expressed as:

$$r \approx \frac{n \cdot a}{2\pi} \quad (5)$$

$$l \approx mb \quad (6)$$

The total energy of nanotube is then expressed as:

$$E_n \approx mn \cdot E_{unit} + 2\frac{\pi r^2}{ab} E_{unit} + \Delta E \approx \left(mn + \frac{n^2 a}{2\pi b}\right) E_{unit} + \Delta E \quad (7)$$

where $E_{unit}$ represents the energy per primitive cell of the corresponding nanosheet; and the additional term $\Delta E$ accounts for other energy components to form a closed nanotube through curling and sealing of nanosheet, which including: (i) bond angle bending energy due to nanosheet curling; (ii) strain energy induced by structural deformation arising from sealing the side-terminals; and (iii) binding energy associated with atomic permeation of the encapsulated matter through the nano-envelope. These factors depend on the detailed structure and the type and phase of the encapsulated materials, and they will always lower the containing pressure, thus we can ignore them here since our main purpose is to estimate the absolute limits of an ideal nanotube/nanosphere. Another argument to ignore this term is that the energy variation with volume $\frac{\delta(\Delta E)}{\delta V}$ should be small if the structure keeps almost unchanged during the dilation/shrinking process. In this way, the energy change with respect to the radius $r$ and length $l$ at the condition that the total number of primitive cells being fixed are:

$$\frac{\delta E_n}{\delta r} \approx \frac{2\pi mb + na}{b} \cdot \frac{\delta E_{unit}}{\delta a} \tag{8}$$

$$\frac{\delta E_n}{\delta l} \approx \frac{2\pi mnb + n^2 a}{2\pi mb} \cdot \frac{\delta E_{unit}}{\delta b} \tag{9}$$

Thus, the utmost limit pressure $P$ can be expressed as:

$$P \approx \frac{\delta E_n}{\delta r} \cdot \frac{\delta r}{\delta V} + \frac{\delta E_n}{\delta l} \cdot \frac{\delta l}{\delta V}$$

$$= \frac{2\pi mb + na}{mnab^2} \cdot \frac{\delta E_{unit}}{\delta a} + \frac{4\pi mb + 2na}{mna^2 b} \cdot \frac{\delta E_{unit}}{\delta b} \tag{10}$$

For simplicity, we approximate Eq. (10) further by taking the primitive cell counts along radial and axial directions are equivalent (i.e., $m = n$). In this special case, the pressure $P$ can be written as:

$$P \approx \frac{1}{r}\left(\frac{2\pi b + a}{2\pi b^2} \cdot \frac{\delta E_{unit}}{\delta a} + \frac{2\pi b + a}{\pi ab} \cdot \frac{\delta E_{unit}}{\delta b}\right)$$

$$= \frac{1}{r}\left(k_1 \cdot \frac{\delta E_{unit}}{\delta a} + k_2 \cdot \frac{\delta E_{unit}}{\delta b}\right) \tag{11}$$

where the structure factors $k_1 = \frac{2\pi b + a}{2\pi b^2}$ and $k_2 = \frac{2\pi b + a}{\pi ab}$. They satisfy $k_1 = \frac{a}{2b} k_2$. Eq. (10) establishes a relation between the cavity size of a nanotube (for $m = n$ case) and the utmost pressure it can preserve, with the coefficients $\frac{\delta E_{unit}}{\delta a}$ and $\frac{\delta E_{unit}}{\delta b}$ can be easily computed using the primitive cell of the nanosheet with first-principles method.

**(A.II) Monolayer nanosphere**

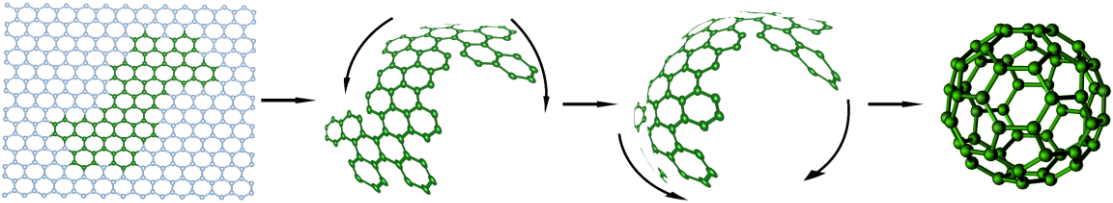

**Fig. 2** (color online) Schematic illustration of graphene wrapped into buckyball with a nano-cavity.

Besides nanotube, graphene as a 2D building block of carbon materials[33], it also can be wrapped into 0D fullerenes, as illustrated in Fig. 2. Here, we take the fullerene as an example to estimate the pressure-bearing capability of a nanosphere. Assuming that the fullerene has a radius of $r_f$ and a surface area of $S_f$, similar to the model presented in (A.I), the total energy can be described as:

$$E = E_m + E_f \tag{12}$$

$$E_f = \epsilon_f \cdot S_f = \epsilon_f \cdot 4\pi r_f^2 \tag{13}$$

where $E_f$ represents the energy of the nanosphere, and $\epsilon_f$ denotes the energy density of the nanosphere per unit area. Let us assume that the nanosphere consists of $n$ primitive cells wrapped from the nanosheet, by ignoring the pentagonal topological defects and curvature-induced bond distortion effects of wrapping and sealing of the nanosphere, we have:

$$S_f \approx S_n = nab = 4\pi r_f^2 \tag{14}$$

$$E_f \approx n \cdot E_{unit} \tag{15}$$

where $S_n$ represents the total area of $n$ primitive cells with lattice constants $a$ and $b$ in a 2D nanosheet, and $E_{unit}$ denotes the energy per primitive cell. The utmost limit pressure $P_f$ of a nanosphere can be expressed at the condition that the total number of primitive cells $n$ being kept fixed when performing the energy variation as:

$$\begin{aligned} P_f &\approx \frac{\delta E_f}{\delta V_f} = \frac{\delta E_f}{\delta r_f} \cdot \frac{\delta r_f}{\delta V_f} \\ &= \left( \frac{\delta E_f}{\delta a} \cdot \frac{\delta a}{\delta r_f} + \frac{\delta E_f}{\delta b} \cdot \frac{\delta b}{\delta r_f} \right) \cdot \frac{\delta r_f}{\delta V_f} \\ &= \frac{2}{r_f} \left( \frac{1}{b} \cdot \frac{\delta E_{unit}}{\delta a} + \frac{1}{a} \cdot \frac{\delta E_{unit}}{\delta b} \right) \end{aligned} \tag{16}$$

**(A.III) Multilayer nanosphere**

The absolute limit of pressure preserving for a multilayer nanosphere can be derived analogously. The total energy $E_F$ of a nanosphere comprising $j$ concentric atomic layers can be expressed as:

$$E_F = E_{f1} + E_{f2} + E_{f3} + \cdots + E_{fj} \tag{17}$$

$$E_{fj} \approx n \cdot E_{unit} + E_{j-1,j} \tag{18}$$

where $E_{fj}$ represents the energy of the $jth$ layer nano envelope, which also includes the interlayer interactions energy with its adjacent layers ($E_{j-1,j}$). Here, compared to the total energy of a unit cell within a single layer (typically in eV/atom), the interlayer interaction energy (typically in meV/atom) exhibits a difference of 1–2 orders of magnitude. Thus, it can be considered negligible. Then the pressure $P_F$ that can be retrieved by such configuration is:

$$P_F \approx \frac{\delta E_F}{\delta V_F} = \frac{\delta E_{f1}}{\delta V_{f1}} + \frac{\delta E_{f2}}{\delta V_{f2}} + \cdots \frac{\delta E_{fj}}{\delta V_{fj}}$$

$$= \left(\frac{2}{r_{f1}} + \frac{2}{r_{f2}} + \cdots \frac{2}{r_{fj}}\right) \cdot \left(\frac{1}{b} \cdot \frac{\delta E_{unit}}{\delta a} + \frac{1}{a} \cdot \frac{\delta E_{unit}}{\delta b}\right)$$

$$= 2 \cdot \left(\sum_{i=1}^{j} \frac{1}{r_i}\right) \cdot \left(\frac{1}{b} \cdot \frac{\delta E_{unit}}{\delta a} + \frac{1}{a} \cdot \frac{\delta E_{unit}}{\delta b}\right) \tag{19}$$

The radius distance between adjacent layers follows a relationship of $r_i = r_{i-1} + \Delta r$. If we assume the interlayer spacing $\Delta r$ between nanosheets is constant, then:

$$r_2 = r_1 + \Delta r \tag{20}$$

$$r_j = r_1 + (j-1) \cdot \Delta r \tag{21}$$

With these simple physical models (Eqs. (11, 16, 19)), we are now on a position to estimate the absolute pressure limit that can be stored by nanotubes or nanospheres. The required key factor is the maximal variation of the total energy per primitive cell with respect to the lattice constants. The involved assumptions are summarized as

follows:

(I) The total internal energy of the nanotube/nanosphere is assumed to be the internal energy of a primitive cell multiplied by the number of primitive cells. Namely, the energy change induced by cell mismatch, pentagonal topological defects, bond bending or distortion are ignored.

(II) The nanotube/nanosphere is assumed having an ideal cylindrical/spherical shape.

(III) The material to be retrieved is uniformly distributed within the nanocavity.

(IV) Interaction between adjacent nano envelopes, and that between the nano-envelope and the contained material, are ignored, as well as the possible atomic permeation through the pores of the nanotube/nanosphere. We note that all of these contributions will reduce the pressure that can be preserved, and the results of models (Eqs. (11, 16, 19)) provide a theoretical estimate of the absolute upper limit.

## B. First-principles computational details

In order to get the maximal $\frac{\delta E_{unit}}{\delta a}$ and $\frac{\delta E_{unit}}{\delta b}$, the total energy calculation and structural optimization of a 2D nanosheet are computed using first-principles methods within the framework of Density Functional Theory[34,35] (DFT) and the Projector Augmented Wave[36] (PAW) method as implemented in the Vienna Ab initio Simulation Package[37] (VASP). The Perdew-Burke-Ernzerhof[38](PBE) functional was adopted for the exchange-correlation functional. To ensure the accuracy of the calculations, the cutoff energy was set to 650 eV, the Gaussian smearing was set to 0.1 eV, and the first irreducible Brillouin zone sampling was performed using a fine enough $15\times15\times1$ Monkhorst-Pack grid. The structural optimization was carried out with an energy tolerance of $10^{-7}$ eV/atom and a force tolerance of -0.001 eV/Å. The k-point convergence tests for the total energies of the metastable states are included in the supplementary materials. For the multi-walled nanotube models, the energy of the system was corrected by using the optB88-vdW functional to account for the van der Waals interactions between the nano layers. The electron/hole doping and chemical adsorption effects were mimicked using the jellium model, namely, to adjust the position of Fermi level by varying the number of electrons in the system with the NELECT tag in VASP. This allows us to assess the impact of adding/removing electrons to the nanotube/nanosphere on their pressure-bearing capability. In this method, the excessive charges are automatically neutralized by uniform background charges.

## III. Results and Discussion

### A. Monolayer nanotube

As described by the theoretical model established above, the capability of nanotube/nanosphere to preserve high-pressure is determined by the maximal $\frac{\delta E_{unit}}{\delta a}$ and $\frac{\delta E_{unit}}{\delta b}$ for a given geometry configuration. We systematically explored and screened the pressure-bearing capability of all possible candidates, here only four representative 2D nanomaterials that are most promising are given: graphene, h-BN, biphenylene, and $\gamma$-graphyne. To facilitate the first-principles computation, all structures were aligned to an orthogonal unit cell, as shown in Fig. 3. The redefined orthorhombic lattice parameters of graphene are $a$=2.46 Å and $b$=4.25 Å, with a C-C bond length of 1.418 Å. For h-BN (and biphenylene), the orthorhombic lattice parameters are $a$=2.51 Å (3.76 Å) and $b$=4.35 Å (4.52 Å), respectively, with B-N bond lengths ranging from 1.45 to 1.48 Å (C-C bond lengths from 1.40 to 1.46 Å), respectively. As for $\gamma$-graphyne, which has a unique monoclinic structure, we used a redefined lattice to convert the monoclinic system into an orthogonal one. The redefined lattice parameters are $a$ = 11.95 Å and $b$ = 6.9 Å, with C-C bond lengths ranging from 1.22 Å to 1.43 Å. To ensure the accuracy of the DFT calculations, a 20 Å vacuum layer was adopted in the z-direction for all structures, to eliminate the interlayer interactions that may arise from periodic boundary conditions.

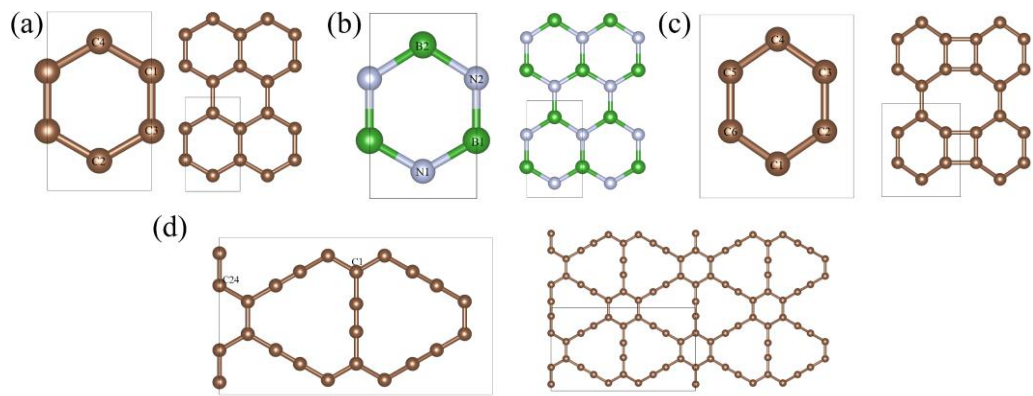

**Fig. 3** (color online) Lattice structures of the orthorhombic primitive cells and 2×2 supercells of four selected 2D materials: (a) graphene, (b) h-BN, (c) biphenylene, and (d) $\gamma$-graphyne.

In order to evaluate the pressure-preserving performance of the four nano-materials, we first calculated the energy change rate under uniaxial tension, namely, $\frac{\delta E_{unit}}{\delta a}$ and $\frac{\delta E_{unit}}{\delta b}$, for 1×1, 2×2, and 3×3 supercells, respectively. As shown in Fig. 4, the results reveal that as the supercell size increases, the peak of the energy change rate per primitive cell for all systems remains nearly constant, as was expected, showing there is no stress localization in these systems. Additionally, the peak in $a$ direction is much higher than along $b$ direction for graphene, h-BN, and biphenylene, suggesting that higher pressure-bearing capability can be obtained when rolling them into nanotubes along the lattice direction $a$. In contrast, $\gamma$-graphyne nanotubes display a higher pressure-bearing capability when rolled along the lattice direction $b$. Using these DFT results and Eq. (11), the pressure-bearing capability of these four 2D materials are calculated and ranked from highest to lowest as follows: graphene ($P \approx \frac{1}{r}145$ GPa), h-BN ($P \approx \frac{1}{r}124$ GPa), biphenylene ($P \approx \frac{1}{r}110$ GPa), and $\gamma$-graphyne ($P \approx \frac{1}{r}108$ GPa), respectively. Here, the nanotube radius $r$ is in a unit of nm. It suggests that graphene is the best candidate for containing high-pressure materials.

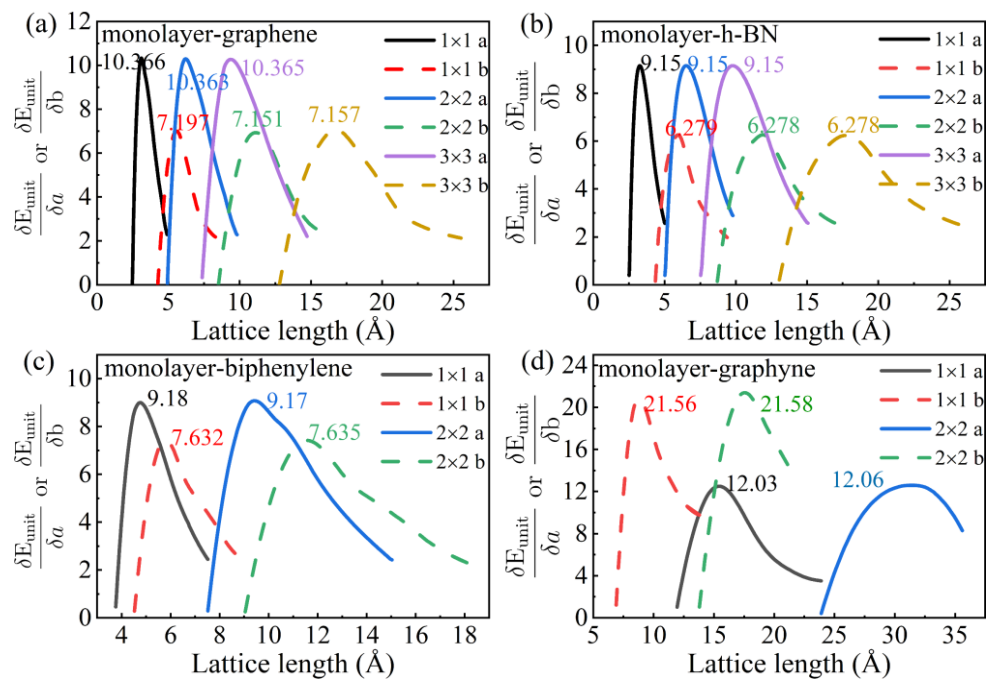

**Fig. 4** (color online) Calculated energy change rates under uniaxial tension for monolayer nanomaterials. (a) graphene, (b) h-BN, (c) biphenylene, and (d) $\gamma$-graphyne. Note the peak value determines the absolute limit.

We also performed calculations using the HSE06 hybrid functional to calculate the energy change rate under uniaxial tension for graphene and hexagonal boron nitride within a 1×1 unit cell. The results are compared with those obtained using the PBE functional, as shown in Fig. S1 (see supplementary materials). Our findings reveal that the HSE06 hybrid functional exhibits improved accuracy in total energy calculations compared to PBE, with higher peak values of the energy change rate. The maximum pressure-bearing capabilities of graphene and hexagonal boron nitride increased from ($P \approx \frac{1}{r}$ 145 GPa/ $P \approx \frac{1}{r}$ 124 GPa, PBE) to ($P \approx \frac{1}{r}$ 151 GPa/ $P \approx \frac{1}{r}$ 133 GPa, HSE06), respectively. This improvement confirms that with increased computational precision, the maximum pressure-bearing capability of materials can be further optimized and enhanced by 3% ~ 6%, while the relative ranking of their pressure-bearing capabilities remains unchanged.

## B. Influence of inter-layer interaction for multilayer nanomaterial

It is helpful if we can quantify the influence of interlayer interactions on their pressure-bearing capability. For this purpose, we consider a bilayer nanotube, and calculate the energy change rate of the bilayer nanotubes with respect to uniaxial tension. In order to describe the interlayer van der Waals interactions accurately, the optB88-vdw *xc* functional was employed. The interlayer spacings $\Delta r$ of 3.42 Å, 3.32 Å, 3.51 Å, and 3.45 Å were adopted for graphene, h-BN, biphenylene, and $\gamma$-graphyne, respectively. The results are shown in Fig. 5, which is evident that the peak of the energy change rate of a bilayer nanotube is approximately twice that of the monolayer structure, indicating that interlayer interaction does not affect the tensile strength of each monolayer nanosheet. This is because although interlayer attractive forces are prevalent in layered materials, such interactions primarily influence interlayer compression, slip, and shear behavior, rather than in-plane tensile stiffness, which is the latter being the primary determinant of the pressure-bearing capability in nanomaterials.

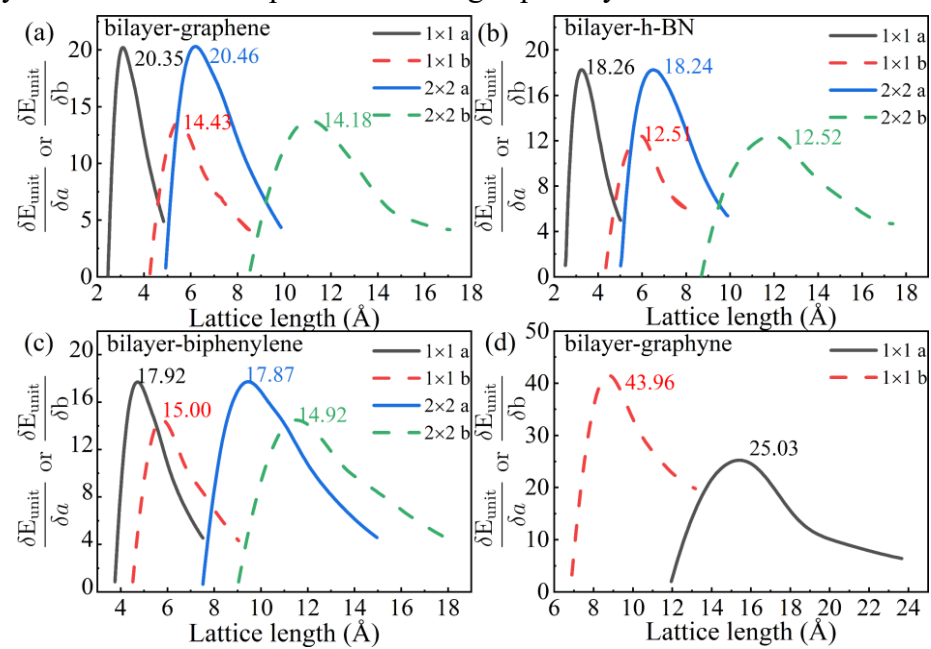

**Fig. 5** (color online) Calculated energy change rates under uniaxial tension for bilayer nanomaterials. (a) graphene, (b) h-BN, (c) biphenylene, and (d) $\gamma$-graphyne.

This finding supports the multilayer model of Eq. (19). That is, the total pressure that can be preserved by a multilayer nanostructure is the linear superposition of the capability of each layer. Therefore, we can enhance the capability of a nanostructure to bear higher pressure by adding more layers to it, which could be an effective method for ultra-high pressure storge.

Furthermore, we constructed and calculated the energy change rate for four slip models of bilayer graphene, with layer shifts of 0.25, 0.5, and 0.75 lattice periods along the a-direction, as shown in Figs. S2 and S3. Our results show that incorporating slip effects moderately lowers the peak energy change rate in the bilayer structure, suggesting a pressure-bearing capability that is slightly reduced compared to the current model predictions. This essentially reflects the intrinsic stiffness of monolayer material. Thus, while interlayer slip influences the efficiency of load transfer within multilayer systems, it does not alter the fundamental mechanical properties of the individual monolayer.

During material preparation and pressurization, multilayer nanospheres are highly susceptible to geometric instability, often resulting in the formation of wrinkles and localized stress concentrations, which will cause excessive stretching of covalent bonds and reduce the number of effective bonds within the layer. In such stacked configurations, the contribution of the effective bond energy to the overall mechanical strength will decrease. These effects are not considered in our ideal model at present.

## C. Electron/hole doping and chemical adsorption influence

Above results are for ideal nanomaterials. It is well known that real 2D nanosheets are easy to adsorb other elements, which might change the chemical environment of each bond (as well as their bond strength) through electron/hole doping into the nanosheet. We investigate the influence of this effect by using the jellium model, in which by increasing or decreasing the total number of electrons in the system, we can adjust the position of Fermi level, thus to mimic the chemical environment of electron/hole doping. The results for the monolayer systems are shown in Fig. 6.

For graphene, h-BN and biphenylene, the peak of energy change rates in both *a* and *b* directions slightly decrease with small hole doping, thus will reduce their pressure-preserving capability slightly. Taking graphene as an example, loss of electrons will reduce the electron density on the $\pi$ bonds, decrease their delocalization, and thus weaken the bond strength.

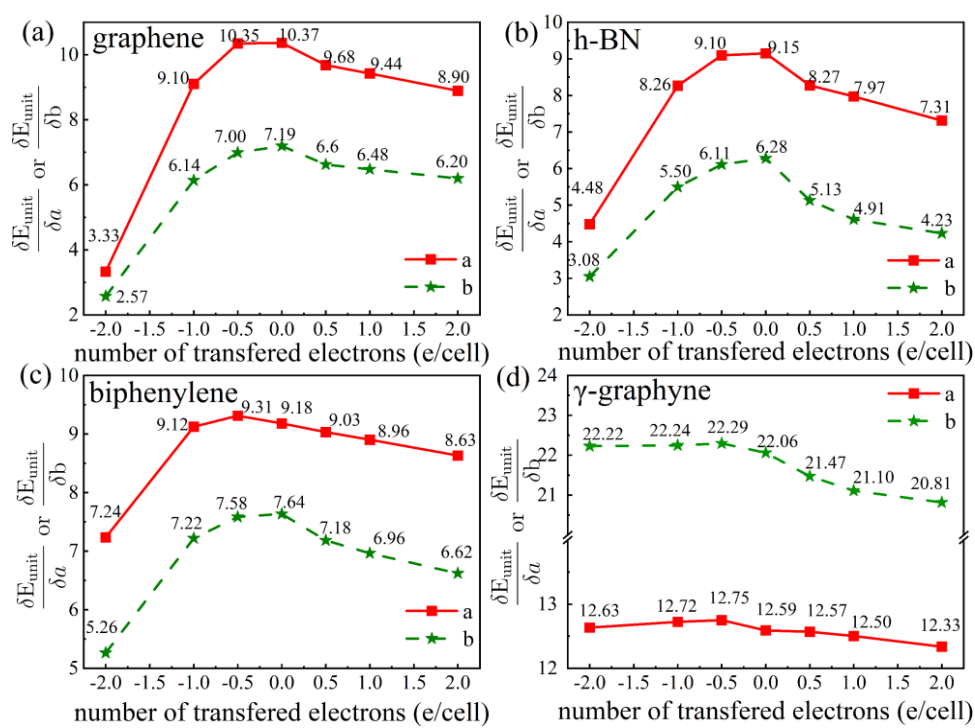

**Fig. 6** (color online) The influence of electron transfer on the peak value of $\frac{\delta E_{unit}}{\delta a}$ and $\frac{\delta E_{unit}}{\delta b}$ for monolayer 2D materials. (a) graphene, (b) h-BN, (c) biphenylene, and (d) $\gamma$-graphyne.

In contrast, for $\gamma$-graphyne, loss of electrons will remove electrons from the antibonding $\pi^*$ orbitals that have a higher energy. This reduces the repulsion between electrons, leading to contraction of the length of some triple-bonds, and enhances the bond strength slightly, as well as the peak value of $\frac{\delta E_{unit}}{\delta a}$ and $\frac{\delta E_{unit}}{\delta b}$. On the other hand, for all structures considered here, adding electrons to the system will increase the electron-electron repulsion, which destabilizes the structure and reduces the pressure-bearing capability. That is to say, chemical doping will not enhance the capability of nanomaterials to contain high pressure generally.

**D. Isotropic biaxial tension**

The uniaxial tension calculation and analysis as mentioned above provide a simplified estimation of the optimal pressure-bearing performance of the selected structures. In order to appraise the absolute limit of various nanostructures to contain high-pressure material more practically, we further calculated their energy change rate under simultaneous biaxial tension since usually the dilation of nanotube/nanosphere is isotropic. The results of $\frac{\delta E_{unit}}{\delta a}$ and $\frac{\delta E_{unit}}{\delta b}$ for these four 2D nanosheets are shown in Fig. 7. Using these DFT results and Eq. (11), the utmost pressures under biaxial tension of four materials are obtained and listed in Table I, respectively.

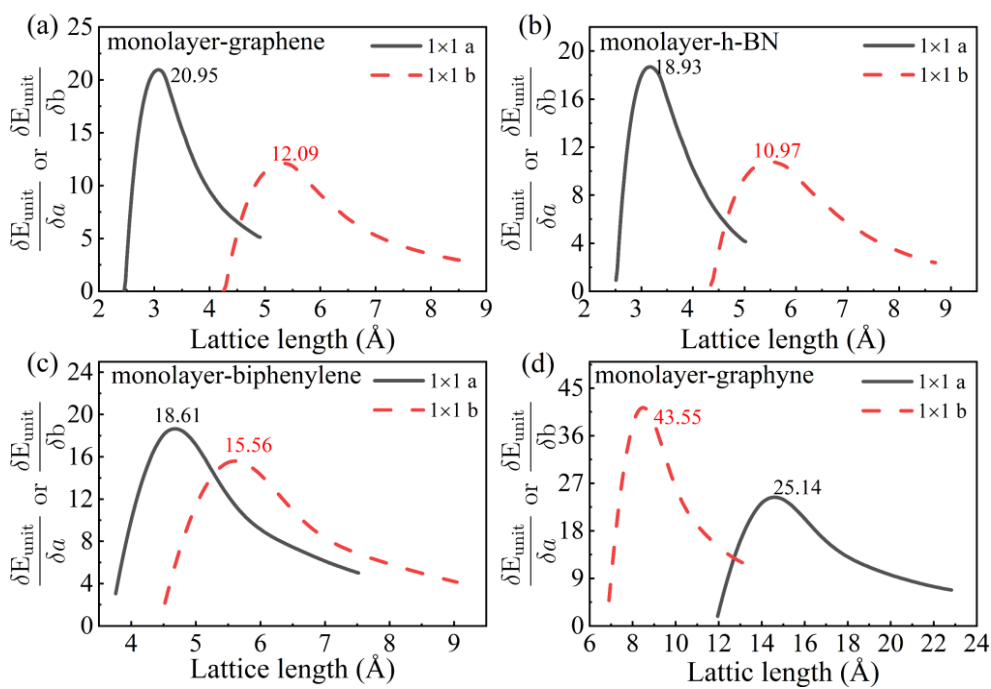

**Fig. 7** (color online) Calculated energy change rates under isotropic biaxial tension for monolayer nanomaterials. (a) graphene, (b) h-BN, (c) biphenylene, and (d) $\gamma$-graphyne.

The results can be understood in terms of average binding energy per bond and bond density of the structure. The energy of each bond[39] is defined as the difference between the ground-state energy of all isolated atoms and the total energy of the system, dividing by the total number of bonds. In this way, the average binding energy per bond is expressed as:

$$E_{binding} = \frac{n \cdot E_{atom} - E_{total}}{N_{bonds}} \quad (22)$$

As shown in Table I, our calculation indicates that graphene exhibits the highest average binding energy per bond (and thus is the strongest), followed by h-BN, biphenylene, and $\gamma$-graphyne. In graphene structure, carbon atoms form a hexagonal lattice with $\sigma$ bonds through sp$^2$ hybridization, while the unhybridized pz orbitals forming a delocalized $\pi$-bond network. The hexagonal symmetry of graphene ensures a uniform distribution of both $\sigma$ and $\pi$ bonds, with a C-C bond length of 1.418 Å. This configuration leads to strong chemical bonding, with an average binding energy per bond of 505 kJ/mol (compared with the C-C bond energy of 497 kJ/mol reported in a theoretical study on graphene structure using DFT-PBE calculations by Shin *et al*.[40]). Despite the structural similarity between h-BN and graphene, the ionic nature of the B-N bond leads to an asymmetric electron distribution in the $\pi$ bonds, causing some degree of electron localization. As a result, the averaged binding energy per bond of h-BN is slightly lower than that of graphene, with 482 kJ/mol per bond. Biphenylene, while having a full sp$^2$-hybridizion carbon network, is a complex structure with four-membered, six-membered and eight-membered rings. This variation in ring sizes leads to uneven bond lengths, with some C-C bonds in the four-membered ring measuring 1.46 Å, exhibiting double-bond characteristics. This variation in bond length weakens the overall bond strength, resulting in an averaged binding energy per bond of 475 kJ/mol. In contrast, $\gamma$-Graphyne has a carbon network with alternating sp and sp$^2$ hybridization. The sp-hybridized orbitals form C≡C triple bonds, while the sp²-hybridized orbitals form C=C double bonds. The alternating triple and double bonds exacerbate the disparity in bond lengths, which range from 1.20 to 1.43 Å. This heterogeneity in bond lengths further weakens the bond energy, resulting an averaged binding energy per bond of 461 kJ/mol.

On the other hand, as demonstrated in Eqs. (11, 16, 19), the bond density (number of bonds per area) also is a key factor to determine the pressure-preserving capability of a nanostructure, because we can rewrite $E_n$ as $E_n = N_b \cdot E_{binding} = \rho_b \cdot S \cdot E_{binding}$, where $N_b$ is the total number of bonds and $S$ is the area of the nanostructure. We also listed in Table I the bond density for these four materials. As can be seen, graphene not only has the strongest bond, but also the highest bond density, suggesting it should have the highest pressure-preserving capability. This observation is consistent with the first-principles results. Furthermore, since graphene has the highest $\rho_b \cdot E_{binding}$ for all known 2D nanomaterials, we can conclude that it is also the best one for high-pressure preservation.

**TABLE I.** Comparison of average binding energy per bond, bond density, and the utmost preserving pressure of a nanotube with monolayer graphene, h-BN, biphenylene, and $\gamma$-graphene under isotropic biaxial tension, respectively. The radius of cavity size of the nanotube $r$ is in a unit of nm.

| Structure | Average binding energy per bond (kJ/mol) | Bond density $\rho_b$(1/ Å³) | $\frac{\delta E_{unit}}{\delta a}$ | $\frac{\delta E_{unit}}{\delta b}$ | Pressure (GPa) |
|---|---|---|---|---|---|
| graphene | 505 | 0.57 | 20.95 | 12.09 | $P \approx \frac{1}{r} 258$ |
| h-BN | 482 | 0.55 | 18.93 | 10.97 | $P \approx \frac{1}{r} 229$ |
| biphenylene | 475 | 0.53 | 18.61 | 15.56 | $P \approx \frac{1}{r} 225$ |
| $\gamma$-graphyne | 461 | 0.44 | 25.14 | 43.55 | $P \approx \frac{1}{r} 223$ |

### E. Monolayer and multilayer nanosphere

Above mentioned calculation and analysis validate the model [Eqs. (11, 16, 19)] to estimate the theoretical upper limit of pressure-preserving capability of 2D nanomaterials, and selected graphene out as the best candidate. In this subsection, we apply this model to nanosphere of graphene to estimate its potential to retrieve interesting novel state of high-pressure materials. We first evaluate the pressure-bearing capability of monolayer nanospheres, by using Eq. (16). The results are ranked from highest to lowest as follows: graphene ($P \approx \frac{1}{r}315$ GPa), h-BN ($P \approx \frac{1}{r}279$ GPa), biphenylene ($P \approx \frac{1}{r}264$ GPa), and $\gamma$-graphyne ($P \approx \frac{1}{r}233$ GPa), respectively. The variation of the preserved pressure as a function of the nanosphere radius (which represents the size of the nanocavity) is shown in Fig. 8(a). Notably, though increasing the nanosphere radius can increase the volume to store the material, it also reduces the utmost pressure that can be preserved. We further calculate the allowed most inner radius of a multilayer nanosphere for the purpose to retrieve a given target pressure as a function of the number of layers. Note the inter-layer spacing $\Delta r$ with a value of 0.342, 0.332, 0.351, and 0.345 nm in Eq. (21) are adopted for multilayer graphene, h-BN, biphenylene, and $\gamma$-graphyne nanospheres calculations, respectively. The results of graphene are plotted in Fig. 8(b). As can be seen, the pressure-bearing capability exhibits a positive correlation with the number of layers. Consequently, to achieve higher pressure-bearing capability with larger material dimensions, multilayer configuration is necessary. For instance, in order to recover a material with a radial size of 8 nm at 150 GPa, a least five graphene layers are required. Figure 8(c) plots the required number of graphene layers to contain a given volume of material (characterized by the most inner radius of the nanosphere) to different pressures, which enabling experimentalists to directly determinate the minimal required number of layers based on the specified pressure and nanocavity volume size.

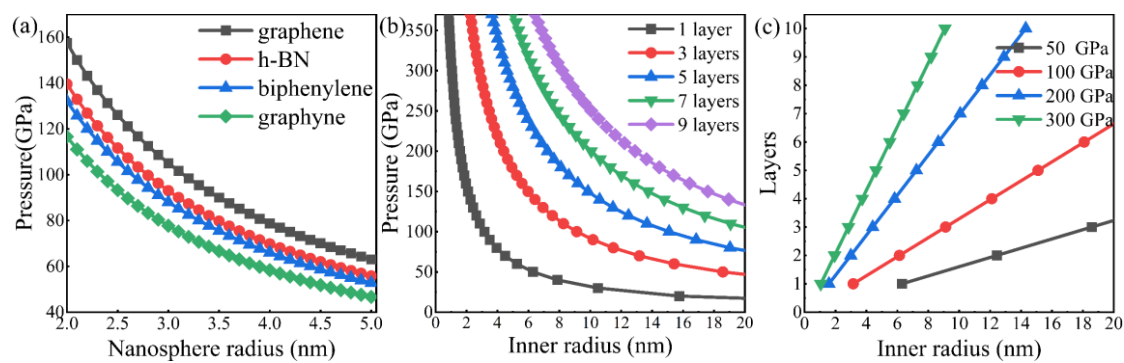

**Fig. 8** (color online) (a) The variation of the utmost preserved pressure for different type of monolayer nanospheres with the radius. (b) The variation of the utmost preserved pressure for a multilayer graphene nanosphere as a function of the most inner radius and number of layers. (c) The variation of the required number of layers for a multilayer graphene nanosphere as a function of the most inner radius to contain different given pressures.

One of important applications of high-pressure preservation is to retrieve room-temperature superconductors. Room-temperature superconductors represent one of the most active research directions in material science and condensed matter physics. The most promising hydrogen-rich compounds that have been experimentally synthesized and theoretically predicted in recent years, along with the molecular metallic hydrogen and atomic metallic hydrogen, are plotted in Fig. 9 (where the *Cmca* phase of molecular metallic hydrogen[41-43], as well as metastable atomic hydrogen at 250 GPa[20,21] with unexplored $T_c$, are presented here with an approximately estimated range for $T_c$ values). In addition, we overlay the curves of the most inner layer radius as a function of pressure that it can be retrieved over the hydrogen-rich compounds data to illustrate the maximum allowable volume of high-pressure materials in different multilayer graphene nanospheres. The maximal volume that can be retrieved by using 1-, 2-, and 3-layer graphene nanospheres to preserve the superconducting state to ambient pressure are calculated for several of these materials and listed in Table II. The results reveal that using only one layer we can effectively retrieve $LaBeH_8$ (191 K at 50 GPa) and $LaBH_8$ (156 K at 55 GPa) to ambient pressure with a maximum volume of more than $10^3$ nm$^3$. Our proposed model Eqs. (11, 16, 19) provides a convenient theoretical guidance for designing a confinement configuration to retrieve high-pressure materials.

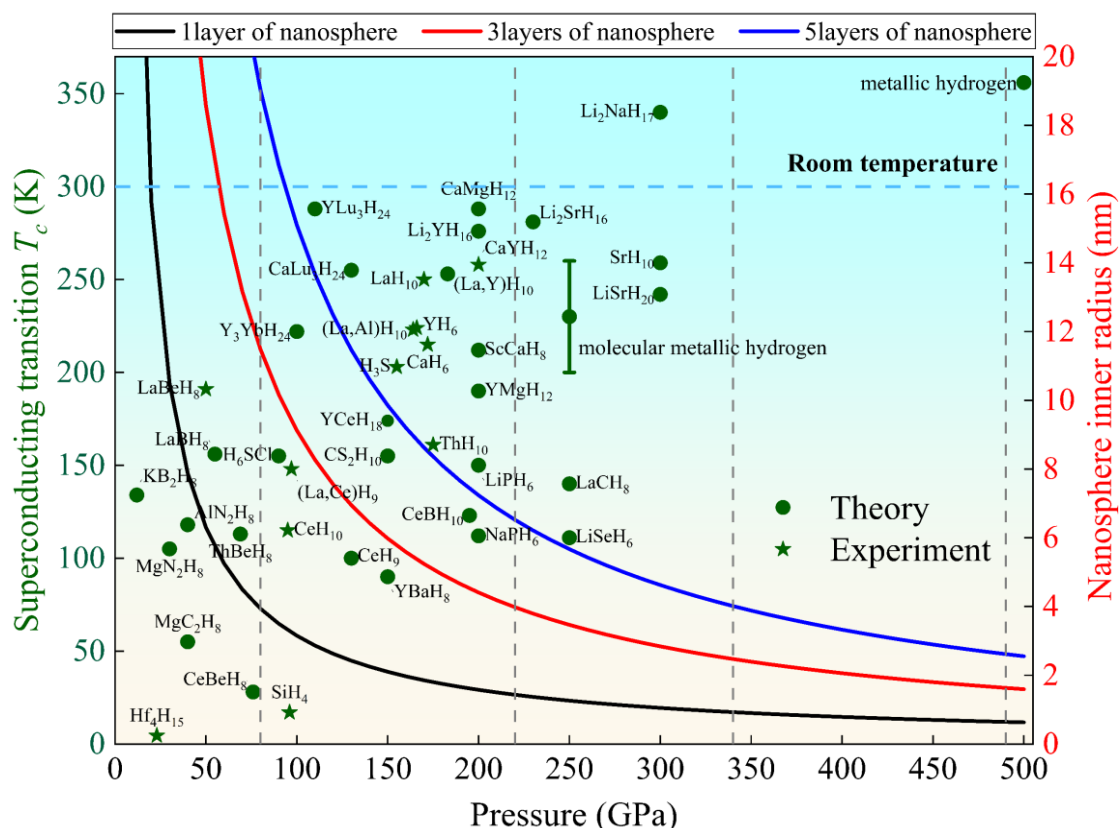

**Fig. 9** (color online) Superconducting materials under high-pressure reported in recent years[11,22-28,44-63]. The green dots and stars correspond to left axis and represent the experimental data and theoretical predictions, respectively. The curves correspond to the right axis. Vertical dashed lines indicate the required number of graphene layers to retrieve them with a given volume (from left to right are one, three, five, and eight layers for a radius of 4 nm of the nano-cavity size, respectively).

**TABLE II.** Number of required layers, retrievable material radius ($r$), and the corresponding cavity volume ($V$) for multilayer graphene nanospheres to preserving different superconducting materials.

| Materials | Pressure (GPa) | $T_c$ (K) | 1 layer | | 2 layers | | 3 layers | |
|---|---|---|---|---|---|---|---|---|
| | | | r (nm) | V (nm$^3$) | r (nm) | V (nm$^3$) | r (nm) | V (nm$^3$) |
| KB$_2$H$_8$[48] | 12 | 134 | 28.25 | 9.4×10$^3$ | 55.86 | 7.3×10$^5$ | 84.90 | 2.56×10$^6$ |
| MgN$_2$H$_8$[50] | 30 | 105 | 10.50 | 4.8×10$^3$ | 20.84 | 3.8×10$^4$ | 31.17 | 1.26×10$^5$ |
| LaBeH$_8$[27] | 50 | 191 | 6.30 | 1.0×10$^3$ | 12.43 | 8.0×10$^3$ | 18.57 | 2.68×10$^4$ |
| LaBH$_8$[28] | 55 | 156 | 5.72 | 440 | 11.35 | 6.1×10$^3$ | 16.88 | 2.0×10$^4$ |
| (La, Ce)H$_9$[11] | 97 | 148 | 3.26 | 145 | 6.35 | 1.0×10$^3$ | 9.43 | 3.5×10$^3$ |
| Y$_3$YbH$_{24}$[52] | 100 | 222 | 3.15 | 130 | 6.13 | 965 | 9.12 | 3.1×10$^3$ |
| YLu$_3$H$_{24}$[52] | 110 | 288 | 2.86 | 98 | 5.56 | 720 | 8.26 | 2.3×10$^3$ |
| CaLu$_3$H$_{24}$[52] | 130 | 255 | 2.42 | 59 | 4.68 | 429 | 6.94 | 1.4×10$^3$ |
| H$_3$S[22,23] | 155 | 203 | 2.03 | 35 | 3.91 | 250 | 5.77 | 804 |
| LaH$_{10}$[24-26] | 170 | 250 | 1.85 | 26 | 3.55 | 187 | 5.23 | 599 |
| molecular metallic H[62] | 250 | 242 | 1.26 | 8.38 | 2.36 | 55 | 3.46 | 173 |
| metallic H[63] | 500 | 356 | 0.63 | 1.05 | 1.11 | 5.73 | 1.59 | 16.84 |

## IV. Conclusion

Theoretical models to evaluate the high-pressure preserving capability of nano-structured cavities by using 2D nano-sheets were established. The absolute limit of various nanotubes/nanospheres for containing high-pressure materials was obtained. Our model reveals that the capability of high-pressure preserving of nanotube/nanosphere is determined by the peak value in the energy change rate with respect to dilation, and can be understood qualitatively by the product of average binding energy per bond and bond density. Using first-principles calculations, we comparatively analyzed four representative 2D nanomaterials: graphene, h-BN, biphenylene, and $\gamma$-graphyne. The results indicate that graphene exhibits the best pressure-bearing capability, followed by h-BN, biphenylene, and $\gamma$-graphyne, respectively. On the other hand, nanospheres show better performance than nanotubes, with the graphene nanosphere having the best capability to preserve high pressure ($P \approx \frac{1}{r}315$ GPa).

We also demonstrated that it could effectively enhance the capability to bear higher pressure by using multilayer configurations of nanosphere. For example, the room-temperature superconductor hydrides $LaH_{10}$ and $YLu_3H_{24}$ can be retrieved to ambient pressure by using a two-layer graphene nanosphere, with a radius size of about 3 and 5 nm, respectively. On the other hand, in order to retrieve the molecular metallic hydrogen or metastable atomic hydrogen (from about 250 GPa) or the atomic metallic hydrogen (from 500 GPa) with a radius size of 4 nm, it requires at least 5 or 8 layers of multilayer graphene nanospheres, respectively. These findings demonstrate the theoretical possibility to retrieve high-pressure materials to ambient conditions by using a nanosphere as the confinement. This insight would stimulate further theoretical and experimental efforts to achieve this goal by developing advanced nano-engineering under high-pressure as well as the techniques to realize self-assembly and to seal the nanospheres at extreme conditions.

## ASSOCIATED CONTENT

### Supporting Information

Additional experimental details and methods, including the computational comparison of graphene and h-BN using PBE and HSE06 functionals, peak energy change rate of different bilayer graphene slip model, and k-point grid convergence tests for orthorhombic γ-graphyne.

## AUTHOR INFORMATION

### Corresponding Authors

**Hao Wang** — *National Key Laboratory of Shock Wave and Detonation Physics, Institute of Fluid Physics, China Academy of Engineering Physics, Mianyang, Sichuan 621900, P. R. China;* https://orcid.org/0000-0001-9125-2091; Email: wh_95@qq.com

**Yanxiao Zhen** — *National Key Laboratory of Shock Wave and Detonation Physics, Institute of Fluid Physics, China Academy of Engineering Physics, Mianyang, Sichuan 621900, P. R. China; School of Science, Key Laboratory of Low Dimensional Quantum Materials and Sensor Devices of Jiangxi Education Institutes, Jiangxi University of Science and Technology, Ganzhou 341000, Jiangxi, P. R. China;* https://orcid.org/0000-0001-6980-0004; Email: yanxiaozhen@jxust.edu.cn

**Huayun Geng** — *National Key Laboratory of Shock Wave and Detonation Physics, Institute of Fluid Physics, China Academy of Engineering Physics, Mianyang, Sichuan 621900, P. R. China; HEDPS, Center for Applied Physics and Technology, and College of Engineering, Peking University, Beijing 100871, P. R. China;* https://orcid.org/0000-0001-5757-3027; Email: s102genghy@caep.cn

## Authors

**Yinli Xu** — *National Key Laboratory of Shock Wave and Detonation Physics, Institute of Fluid Physics, China Academy of Engineering Physics, Mianyang, Sichuan 621900, P. R. China;* https://orcid.org/0009-0004-8127-2629

**Guangfa Yang** — *National Key Laboratory of Shock Wave and Detonation Physics, Institute of Fluid Physics, China Academy of Engineering Physics, Mianyang, Sichuan 621900, P. R. China;* https://orcid.org/0009-0003-1908-5139

**Yi Sun** — *National Key Laboratory of Shock Wave and Detonation Physics, Institute of Fluid Physics, China Academy of Engineering Physics, Mianyang, Sichuan 621900, P. R. China;* https://orcid.org/0009-0001-7923-3557

**Hongxing Song** — *National Key Laboratory of Shock Wave and Detonation Physics, Institute of Fluid Physics, China Academy of Engineering Physics, Mianyang, Sichuan 621900, P. R. China;* https://orcid.org/0009-0001-1901-1207

**Yusong Huang** — *National Key Laboratory of Shock Wave and Detonation Physics, Institute of Fluid Physics, China Academy of Engineering Physics, Mianyang, Sichuan 621900, P. R. China;* https://orcid.org/0009-0007-7422-8242

## Notes

The authors declare no competing financial interest.

## ACKNOWLEDGEMENTS

This work is supported by the National Nature Science Foundation of China under Grant No.12404287 and No.12364003, the Key Laboratory of Low Dimensional Quantum Materials and Sensor Devices of Jiangxi Education Institutes (NO. GanJiaoKeZi-20241301)

## Author Contributions

Yin L. Xu: Calculation, Analysis, Writing, Original draft preparation. Guang F. Yang: Calculation, Analysis, Validation, Editing. Yi Sun: Analysis, Validation. Hong X. Song: Analysis, Validation. Yu Song Huang: Analysis, Validation. Hao Wang: Reviewing and Editing, Fund. Xiao Z. Yan: Writing, Reviewing and Editing, Fund. Hua Y. Geng: Idea conceiving, Project design, Writing, Reviewing and Editing, Fund.

**Table of Contents.** The curves of the most inner layer radius as a function of pressure that it can be retrieved over the hydrogen-rich compounds data to illustrate the maximum allowable volume of high-pressure materials in different multilayer graphene nanospheres. For Table of Contents Only.

# Supplementary Material

# Appraising the absolute limits of nanotubes/nanospheres to preserve high-pressure materials

Yin L. Xu[1,*], Guang F. Yang[1,*], Yi Sun[1], Hong X. Song[1], Yu S. Huang[1], Hao Wang[1,†], Xiao Z. Yan[1,2,†] and Hua Y. Geng[1,3,†]

[1] *National Key Laboratory of Shock Wave and Detonation Physics, Institute of Fluid Physics, China Academy of Engineering Physics, Mianyang, Sichuan 621900, P. R. China;*

[2] *School of Science, Key Laboratory of Low Dimensional Quantum Materials and Sensor Devices of Jiangxi Education Institutes, Jiangxi University of Science and Technology, Ganzhou 341000, Jiangxi, P. R. China;*

[3] *HEDPS, Center for Applied Physics and Technology, and College of Engineering, Peking University, Beijing 100871, P. R. China.*

---

\* *Contributed equally.*
† *Corresponding authors.* E-mail: wh_95@qq.com; yanxiaozhen@jxust.edu.cn; s102genghy@caep.cn

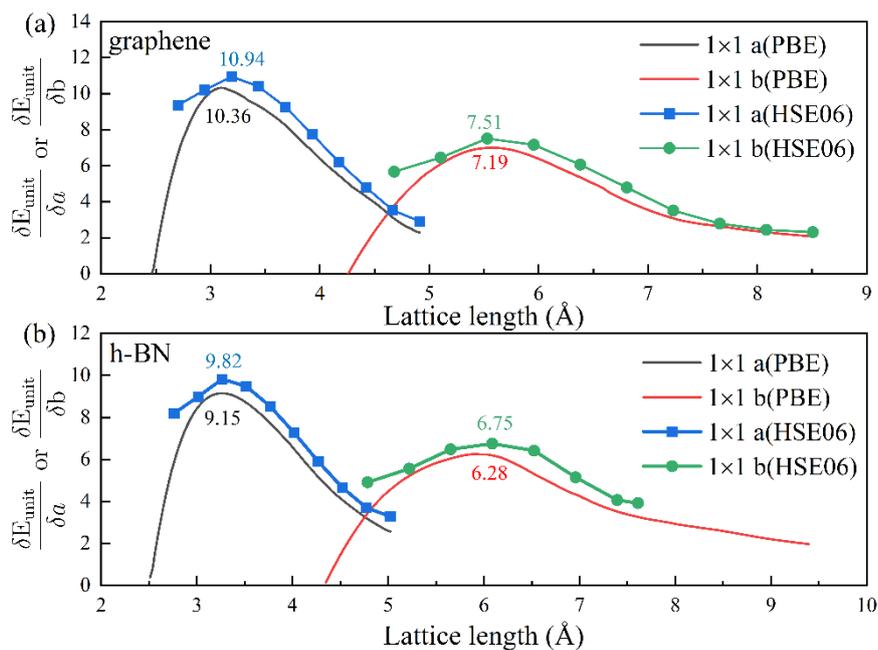

**Fig. S1** (color online) Comparison of energy change rates calculated with different functional under uniaxial tension conditions. (a) graphene, (b) h-BN.

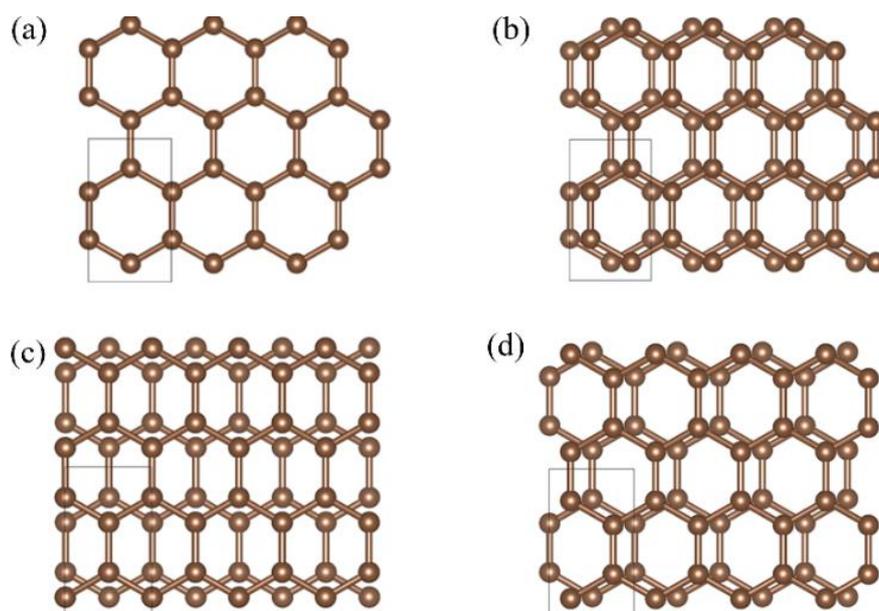

**Fig. S2** (color online) Bilayer graphene slip model. (a) AA stacking, (b) shift of 0.25a, (c) shift of 0.5a, (d) shift of 0.75a. Here, a represents the lattice constant.

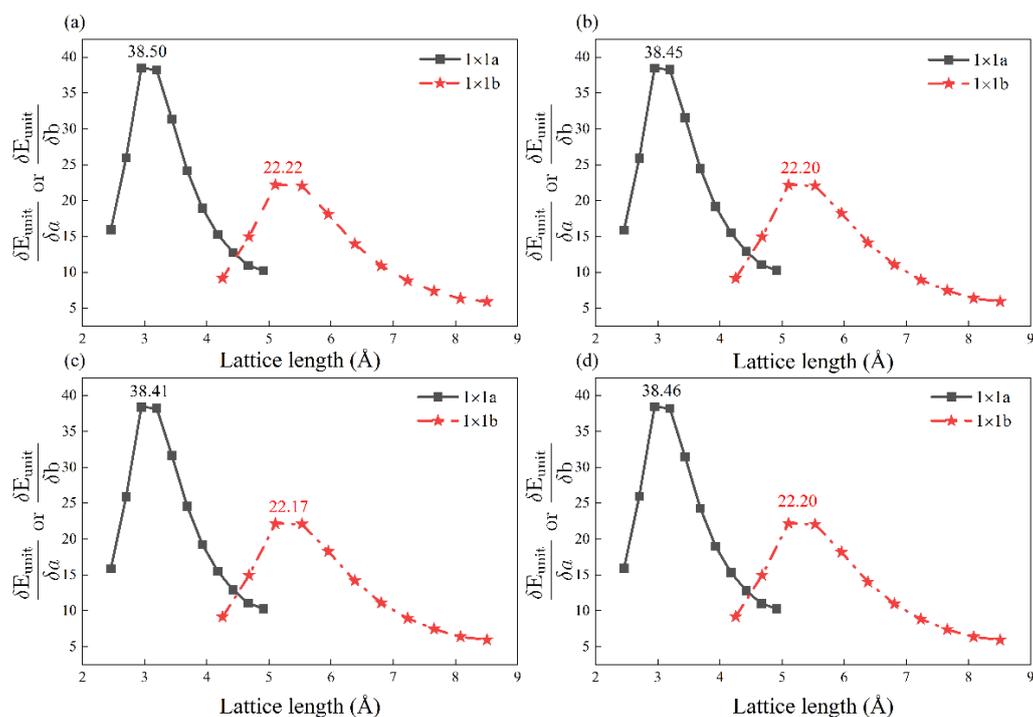

**Fig. S3** (color online) The peak energy change rate of different bilayer graphene slip model. (a) AA stacking, (b) shift of 0.25a, (c) shift of 0.5a, (d) shift of 0.75a.

**TABLE SI**. The k-point grid convergence tests for orthorhombic γ-graphyne.

| K-mesh | Energy(eV) | Delta H(eV/atom) |
|---|---|---|
| 1×1×1 | -206.88396 | |
| 2×2×1 | -205.82603 | 0.04408 |
| 3×3×1 | -206.18434 | -0.01493 |
| 4×4×1 | -206.14276 | 0.00173 |
| 5×5×1 | -206.15920 | -0.00068 |
| 6×6×1 | -206.15625 | 0.00012 |
| 7×7×1 | -206.15648 | -0.00001 |
| 8×8×1 | -206.15765 | -0.00005 |
| 9×9×1 | -206.15733 | 0.00001 |
| 10×10×1 | -206.15840 | -0.00004 |
| 11×11×1 | -206.15803 | 0.00002 |
| 12×12×1 | -206.15833 | -0.00001 |
| 13×13×1 | -206.15821 | 0.00001 |
| 14×14×1 | -206.15814 | 0.00000 |
| 15×15×1 | -206.15812 | 0.00000 |

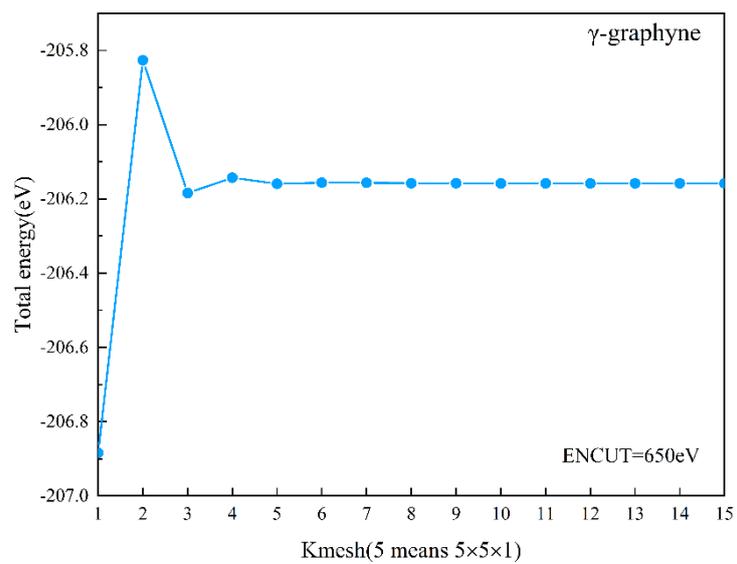

**Fig. S4** The total energy convergence of orthorhombic γ-graphyne with K-mesh (ENCUT=650 eV).